\documentstyle[12pt]{article}

\begin{document}

\title{GEODESICS IN A QUASI--SPHERICAL SPACETIME: A CASE OF GRAVITATIONAL REPULSION}
\author{L. Herrera\thanks{Postal address: Apartado 80793, Caracas 1080 A, Venezuela; e-mail address:
laherrera@telcet.net.ve}
\\
Escuela de F\'\i sica, Facultad de Ciencias,\\
Universidad Central de Venezuela,\\
Caracas, Venezuela.\\
}

\date{}
\maketitle
\newpage
\begin{abstract}
Geodesics are studied in one of the Weyl metrics, referred to as the M--Q solution. First, arguments are
provided, supporting our belief that this space--time is the more suitable (among the known solutions of the Weyl family) for discussing the
properties of strong
quasi--spherical gravitational fields. Then, the behaviour of geodesics is compared with the spherically symmetric situation, bringing out the sensitivity of the trajectories to
deviations from spherical symmetry. Particular attention deserves the change of sign in proper radial acceleration of test particles moving radially along symmetry axis, close to the
$r=2M$ surface, and related to the quadrupole moment of the source.
\end{abstract}
\newpage
\section{Introduction}

As we know, all physical systems in nature (without exception) are submitted to fluctuations, and of course  this well established fact also applies to self--gravitating systems.
Accordingly we have to accept that any physical property of such a  system, e.g. spherical symmetry (if assumed ) is  to be submitted to such fluctuations. In this paper we shall
deal with two obvious questions  arising from these general comments, namely:
\begin{itemize}
\item How to describe such  fluctuations?.
\item  How sensitive are the properties of compact objects to deviations from spherical symmetry?.
\end{itemize}

Now, if the field produced by such self--gravitating systems is not particularly intense (the boundary of the source is
much larger than the horizon) and the deviation from spherical symmetry is
slight, then there is not problem in
representing the corresponding deviations from spherical symmetry (both inside and outside the source)
as a suitable perturbation of the spherically symmetric exact solution \cite{Letelier} (although strictly speaking the term ``horizon'' refers to the spherically symmetric case, we
shall use it when considering $r=2M$ surface, in the case of small deviations from sphericity). 

However, as the object becomes more
and more compact, such perturbative schemes will eventually fail close to the source. Indeed, as is well known \cite{4}, though usually
overlooked, as the boundary surface of the source approaches the horizon (in the sense indicated above), any finite
perturbation of the Schwarzschild spacetime, becomes fundamentally different
from any Weyl metric, even if the latter is characterized by parameters
whose values are arbitrarily close to those corresponding to spherical
symmetry. In other words, for strong gravitational fields, no matter how small are the multipole moments (higher than monopole)
 of the source, there exists a bifurcation between the perturbed Schwarzschild metric  and all the
other Weyl metrics.  This in turn, is just the expression of the fact \cite{1}, that
the
only static and asymptotically-flat vacuum space-time possessing a
regular
horizon is the Schwarzschild solution. For all the others Weyl exterior
solutions \cite{2}, the physical components of the Riemann tensor
exhibit
singularities at $r=2M$.\par\par

In line with these arguments, let us mention that in  recent works \cite{Herrera}, \cite{Herreramass} it has been shown that such bifurcation between the exactly spherically symmetric
case and a solution of the Weyl family, when considering  the source (the interior), takes place for strong gravitational fields, but before the horizon is reached.

 It is worth mentioning that 
the sensitivity of the trajectories of test particles in the
$\gamma$ spacetime \cite{zipo}, to small changes of $\gamma$, for orbits close to $2M$,
has been brought out \cite{HS}. Also, the influence of the quadrupole moment on the motion of test particles within the context of Erez--Rosen metric \cite{erroz} has been investigated
by many authors (see \cite{quevedo1}, \cite{quevedo2} and references therein)

Motivated by the above mentioned facts, we want to present  in this work another approach to the problem of
describing deviations from spherical symmetry. This consists in
considering an exact solution of Einstein equations (of the Weyl family, since we restrict ourselves to static axially--symmetric solutions) continuously linked to
the Schwarzschild metric, through one of its parameters, instead of considering a perturbation of the Schwarzschild space--time.

However, since there are as many
different (physically distinguishable \cite{luisyo}) Weyl solutions as there are different harmonic functions (see
next section), then the  obvious  question arises: which among Weyl solutions is better entitled to describe small deviations from spherical symmetry?.

Although it should be obvious that such question does not have not a unique answer (there is an infinite number of ways of being non--spherical, so to speak), we shall invoke a very
simple criterion, emerging from Newtonian gravity, in order to choose our solution.

Indeed, in order to answer  the question above, let us recall \cite{yo} that most known Weyl solutions,  present a
serious drawback  when describing quasi--spherical space--times. It
consists of the fact that its multipole
structure is such that multipole moments, higher than quadrupole, are of
the same order as the quadrupole moment. Instead, as it is intuitively clear, the
relevance of such multipole moments
should decrease as we move from lower to higher moments, the quadrupole
moment being the most relevant for a small departure from sphericity. Thus
for example in Newtonian gravity,
multipole moments of an ellipsoid of rotation, with homogeneous density,
and axes (a,a,b) read:

\[ \left\{
\begin{array}{lll}
D_{2n} &= &\frac{(-2)^n 3 M a^{2n}}{(2n+1)(2n+3)} \epsilon^n
(1-\epsilon/2)^n ,\qquad \epsilon\equiv \frac{a-b}{a},\\
D_{2n+1} &= & 0
\end{array}
\right. \]

because of the factor $\epsilon^n$, this equation clearly exhibits the
progressive decreasing of the relevance of multipole moments as $n$
increases.

Thus, in order to describe small departures from sphericity, by means of
exact solutions of the Einstein equations, we  would require an exact solution
whose relativistic multipole structure shares the
property mentioned above. Fortunately enough, such a solution exists.

Indeed, there is one (exact) solution  of the Weyl family (hereafter referred to as the M--Q solution \cite{yo}), which may be
interpreted as a quadrupole correction to the Schwarzschild space--time
(see below). It is for
this exterior metric that we are going to study the motion of test particles (more exactly, for a sub--class of this solution).

As we shall see, for very strong fields,  arbitrarily small values of the quadrupole moment of the source, introduce important differences in the radial motion of test particles, as
compared with the exactly spherically symmetric case. Particularly interesting is the appearance of repulsive forces, related to the quadrupole moment, acting on radially moving
particles, along the symmetry axis and close to the horizon.

The paper is organized as follows. In the next section we give a brief resume of Weyl solutions. Next, we describe
the M--Q solution \cite{yo} and comment on 
its properties, in Section 3. The geodesic equations for the M--Q solution are found and analyzed in Section 4. Finally results are discussed in the last
Section.

\section{The Weyl metrics}

Static axysymmetric solutions to Einstein equations are given by the Weyl
metric \cite{2}
\begin{equation}
ds^2 = e^{2 \Psi} dt^2 - e^{-2 \Psi} [e^{2 \Gamma}(d \rho^2 +dz^2)+\rho^2
d \phi^2 ],
\label{elin}
\end{equation}
where metric functions have to satisfy
\begin{equation}
\Psi_{, \rho \rho}+\rho^{-1} \Psi_{, \rho}+\Psi_{, zz} = 0,
\label{meq1}
\end{equation}
and
\begin{equation}
\Gamma_{, \rho}= \rho (\Psi_{, \rho}^2-\Psi_{, z}^2) \quad; \quad \Gamma_{,
z}= 2 \rho \Psi_{, \rho} \Psi_{, z}.
\label{meq2}
\end{equation}

Observe that  (\ref{meq1}) is just the Laplace equation for $\Psi$ (in the
Euclidean space), and furthermore it represents the integrability condition
for (\ref{meq2}), implying that
for any ``Newtonian'' potential we have a specific Weyl metric, a well known result.

The general solution of the Laplace equation (\ref{meq1}) for the function
$\Psi$, presenting an asymptotically flat behaviour, results to be
\begin{equation}
\Psi = \sum_{n=0}^{\infty} \frac{a_n}{r^{n+1}} P_n(\cos \theta),
\label{psi}
\end{equation}
where $r=(\rho^2+z^2)^{1/2}$, $\cos \theta= z/r$ are Weyl spherical
coordinates and $P_n(\cos \theta)$ are Legendre Polynomyals. The
coefficients $a_n$ are arbitrary real constants
 which have been named in the literature ``Weyl moments'', although they
cannot be identified as relativistic multipole moments in spite of the
formal similarity between expression
(\ref{psi}) and the Newtonian potential.

Another interesting way of writting the solution (\ref{psi}) was obtained
by Erez-Rosen \cite{erroz} and Quevedo \cite{quev}, integrating equations
(\ref{meq1}, \ref{meq2})
in prolate spheroidal coordinates, which are defined as follows
\begin{eqnarray}
x & = & \frac {r_{+}+r_{-}}{2 \sigma} \quad , \quad y  =  \frac
{r_{+}-r_{-}}{2 \sigma}\nonumber \\
r_{\pm} & \equiv & [\rho^2+(z\pm \sigma)^2]^{1/2} \nonumber \\
x & \geq & 1 \quad , \quad -1 \leq y \leq 1 ,
\label{pro}
\end{eqnarray}
where $\sigma$ is an arbitrary constant which will be identified later with
the Schwarzschild's mass. 
The prolate coordinate $x$ represents a radial coordinate, whereas the
other coordinate, $y$ represents the cosine function of the polar angle.

In these prolate spheroidal coordinates, $\Psi$ takes the form
\begin{equation}
\Psi = \sum_{n=0}^{\infty} (-1)^{n+1} q_n Q_n(x) P_n(y)
\label{propsi}
\end{equation}
being $Q_n(y)$ Legendre functions of second kind and $q_n$ a set of
arbitrary constants.
The corresponding expression for the function $\Gamma$, has been obtained
by Quevedo \cite{quev}

A sub--family of Weyl solutions has been obtained by Gutsunaev and Manko
\cite{GM85},\cite{M89}
starting from the Schwarzchild solution as a ``seed'' solution.

Both sets of coefficients, ${a_n}$ and ${q_n}$, characterize any Weyl
metric \cite{quev}. Nevertheless, as mentioned before, these constants do
not give us physical information about
the metric since they  do not represent the ``real'' multipole moments of
the source. That is not the case for the relativistic multipole moments
firstly defined by Geroch \cite{ger},
and subsequently by Hansen
\cite{han} and Thorne
\cite{th}, which, as it is known, characterize completely and uniquely, at
least in the neighbourhood of infinity, every asymptotycally flat and
stationary vacuum solution
\cite{kun1}, providing at the same time a physical description
of the corresponding solution.

An algorithm to calculate the Geroch multipole moments was developed by
G.Fodor, C. Hoenselaers and Z. Perjes \cite{fhp} (FHP). By applying such a
method, the resulting multipole
moments of the solution are expressed in terms of the Weyl moments. Similar
results are obtained from Thorne's definition, using harmonic
coordinates.  The structure of the
obtained relation between coefficients $a_n$ and these relativistic moments
allows to express the Weyl moments as a  combination of the Geroch
relativistic moments.

\vskip 1cm

\subsection {The Monopole--Quadrupole solution, $M-Q$}

\vskip 2mm

In this section we would like to review the properties of a solution
(hereafter referred to as the $M-Q$ solution \cite{yo}) which is
particularly suitable for the study of
perturbations of spherical symmetry. The main argument to support this
statement is based on the fact that previously known Weyl metrics (e.g.
Gutsunaev--Manko \cite{GM85}, Manko
\cite{M89}, $\gamma$--metric \cite{zipo}, etc.) have a
multipolar structure (in the Geroch sense) such that all the moments higher
than the quadrupole, are of the same
order as the quadrupole moment. In fact for the above mentioned metrics we have (odd
moments are of course vanishing)
\begin{eqnarray}
 M_0^{GM} & = & M_0^{ER} = M \nonumber\\
 M_2^{GM} & = & M_2^{ER} = \frac{2}{15} q_2 M^3 \nonumber\\
 M_4^{GM} & = & -3 M_4^{ER} = \frac{4}{35} q_2 M^5 \\
 M_6^{GM} & = & M_6^{ER}-\frac 27 \frac{817}{33} M^2 M_4^{ER} =
\frac{2}{15} \frac{4}{231} q_2 M^7 (\frac{194}{7}
+\frac{14}{15} q_2)  \quad ,\nonumber
\label{(14)}
\end{eqnarray}
where $q_2$ is the quadrupole parameter in the  Erez-Rosen metric. For the
gamma metric the results are
\begin{eqnarray}
M_0 &=& \gamma M \nonumber\\
M_2 &=&\gamma \frac{M^3}{3} (1-\gamma^2) \nonumber \\
M_4 &=&\gamma \frac{M^5}{5} (1-\gamma^2) (1-\frac{19}{21} \gamma^2) \nonumber\\
M_6 &=&\gamma \frac{M^7}{7} (1-\gamma^2)
(1+\frac{389}{495}\gamma^4-\frac{292}{165} \gamma^2) \nonumber
\label{MMgamma}
\end{eqnarray}
For the Curzon metric \cite{curzon}, the situation is still worse since
it posseses a unique parameter which represents the mass, and all higher
moments are proportional to increasing powers of that parameter, i.e.,
$
M_0= -a_0,
M_2=\frac13 a_0^3,
M_4=-\frac{19}{105} a_0^5,
M_6= \frac{389}{3465} a_0^7$.

From the comments above it should be clear that neither of the mentioned metrics satisfies the criterion described in the Introduction.

In \cite{yo} it was shown that it is possible to find a solution of the
Weyl family, by a convenient choice of coefficients
$a_n$, such that the resulting solution possesses only monopole and
quadrupole moments (in the Geroch sense). The obtained solution
($M-Q$)
may be written as series of  Gutsunayev--Manko  \cite{GM85}  and
Erez--Rosen \cite{erroz} solutions, as follows:
\begin{equation}
\Psi_{M-Q}=\Psi_{q^0}+q \Psi_{q^1}+q^2 \Psi_{q^2}+\ldots =
\sum_{\alpha=0}^\infty q^\alpha\Psi_{q^\alpha} \quad ,
\label{(15)}
\end{equation}
where the zeroth order corresponds to the
Schwarzschild solution.
\begin{equation}
\Psi_{q^0}=-\sum_{n=0}^{\infty}\frac
{\lambda^{2n+1}}{2n+1}P_{2n}(\cos\theta) \quad ,
\label{(16)}
\end{equation}
with $\lambda \equiv M/r$ and
it appears that each power in $q$ adds a quadrupole correction to the
spherically symmetric solution. Now, it should be observed that due to the
linearity of Laplace equation,
these corrections give rise to a series of exact solutions. In other
words, the power series of $q$ may be cut at any order, and the partial
summation, up to that order, gives an exact
solution representing a quadrupolar correction to the  Schwarzschild solution.

The simplest way to interpret physically the exact solutions obtained from
the quadrupolar corrections described above, consists in analyzying the
corresponding multipolar structure.
Thus , it can be shown that cutting solution (\ref{(15)}) at some order $\alpha$,
one obtains an exact solution with the following properties:
\begin{itemize}
\item Both, the monopole and the quadrupole moments are non--vanishing:

 $M_0\equiv M$, $M_2 \equiv q M^3$.

\item All the  moments beyond $M_2$, until  order $2(\alpha+1)$  (inclusive) vanish.

\item All moments above $M_{2(\alpha+1)}$ are of order $q^{\alpha+1}$. Therefore,
the solution represents a quadrupolar correction to the Schwarzschild
solution, which is an exact
solution up to the given order.
\end{itemize}
Since we are interested in slight deviations from spherical symmetry, we shall consider the M--Q solution, only up to first order in $q$. Thus, 
the explicit solution up to first order, describing a quadrupolar correction to the monopole, is 
( note a misprint in the equation (13) in \cite{yo2})
\begin{eqnarray}
\Psi_{M-Q}^{(1)} &\equiv & \Psi_{q^0}+q \Psi_{q^1} = \frac12
\ln\left(\frac{x-1}{x+1}\right) + \frac58 q(3y^2-1)\times \nonumber\\
&\times&\left[\left(\frac{3x^2-1}4 -\frac1{3y^2-1}\right)
\ln\left(\frac{x-1}{x+1}\right)
 -\frac{2x}{(x^2-y^2)(3y^2-1)} + \frac{3x}2\right] \quad ,
\label{(25)}
\end{eqnarray}

\begin{eqnarray}
\Gamma^{(1)}_{M-Q} & \equiv & \Gamma_{q^0}+q \Gamma_{q^1}+q^2 \Gamma_{q^2} =\frac12\left(1+\frac{225}{24}
q^2\right)\ln\left(\frac{x^2-1}{x^2-y^2}\right) \nonumber\\
&-& \frac{15}8 q x(1-y^2)\left[1- \frac{15}{32}q\left(x^2+7y^2-9x^2y^2+1-\frac83
\frac{x^2+1}{x^2-y^2}\right)\right]\ln\left(\frac{x-1}{x+1}\right) \nonumber\\
&+& \frac{225}{1024}q^2(x^2-1)(1-y^2)(x^2+y^2-9x^2y^2-1)
\ln^2\left(\frac{x-1}{x+1}\right) \nonumber\\
&-& \frac{15}4 q(1-y^2)\left[1-\frac{15}{64}q(x^2+4y^2-9x^2y^2+4)\right]
\nonumber\\
&-& \frac{75}{16}q^2 x^2\frac{1-y^2}{x^2-y^2} - \frac{5}{4}q(x^2+y^2)
\frac{1-y^2}{(x^2-y^2)^2} \nonumber\\
&-& \frac{75}{192}q^2(2x^6-x^4+3x^4y^2-6x^2y^2+4x^2y^4-y^4-y^6)
\frac{1-y^2}{(x^2-y^2)^4}  \quad .
\label{(26)}
\end{eqnarray}

The first twelve Geroch moments of this solution are
(odd moments vanish, because of the reflection symmetry)
\begin{eqnarray}
M_0 &=& M , M_2 = M^3 q , M_4 = 0 ,
M_6 = -\frac{60}{77} M^7 q^2 \nonumber\\
\vspace{0.3cm}
M_8 &=& -\frac{1060}{3003} M^9 q^2 - \frac{40}{143} M^9 q^3 ,
M_{10} = -\frac{19880}{138567} M^{11} q^2 +
\frac{146500}{323323} M^{11} q^3 \nonumber\\
\vspace{0.3cm}
M_{12}&=& -\frac{23600}{437437}M^{13}q^2 + \frac{517600}{1062347}M^{13}q^3 +
\frac{4259400}{7436429}M^{13}q^4
\label{(27)}
\end{eqnarray}
From the expressions above, it is apparent that the parameter
$q\equiv M_2/ M^3$ representing the quadrupole moment, enters into the
multipole moments $M_{2n}$, for $n\ge 2$,
only at order $2$ or higher. Accordingly,  solution (\ref{(25)}, \ref{(26)}) for a small value
of $q$, up to order $q$, may be interpreted as the gravitational field
outside a quasi--spherical source.
The spacetime being represented by a quadrupole correction to the monopole
(Schwarzschild) solution. This is in contrast with other previously
mentioned solutions of Weyl family, where
all moments higher than the quadrupole are of the same order in $q$ as the
quadrupole. And therefore for small values of $q$ they cannot be
interpreted as a quadrupole perturbation of
 spherical symmetry.

Instead of cylindrical corrdinates $(\rho,z)$, it will be useful for the
next section to work with   Erez-Rosen coordinates
$(r,
\theta)$  given by:
\begin{eqnarray}
z &=&(r-M) \cos \theta \nonumber\\
\rho &=& (r^2-2 M r)^{1/2} \sin\theta
\label{(28)}
\end{eqnarray}
and related to prolate coordinates, by
\begin{eqnarray}
x &=& \frac{r}{M}-1 \nonumber\\
y &=& \cos\theta
\label{(29)}
\end{eqnarray}

The metric functions of the  solution, up to the
first order in $q$, hereafter
referred as M-Q$^{(1)}$ are:
\begin{eqnarray}
\Psi_{M-Q}^{(1)}  &  = & \frac12
\ln\left(1-\frac2R\right) + \frac{5}{32} q(3y^2-1)(3 R^2-6
R+2)\ln\left(1-\frac2R\right) \nonumber\\
\vspace{0.2cm}
&-&\frac58 q \ln\left(1-\frac2R\right)-\frac54 q
\frac{R-1}{(R-1)^2-y^2}+\frac{15}{16}q (3 y^2-1) (R-1) \quad ,
\label{(30.1)}
\end{eqnarray}

\begin{eqnarray}
\Gamma^{(1)}_{M-Q} &  = &\frac12
\ln\left[\frac{(R-1)^2-1}{(R-1)^2-y^2}\right]-\frac{15}{8} q
(1-y^2)(R-1)\ln\left(1-\frac2R\right) \nonumber\\
\vspace{0.2cm}
&-&\frac{15}{4} q (1-y^2)-\frac54 q(1-y^2)
\left[\frac{(R-1)^2+y^2}{((R-1)^2-y^2)^2}\right]
\label{(30.2)}
\end{eqnarray}
\vskip 1cm
with $R\equiv r/M$.

It is worth noting that the M--Q solution (\ref{(15)}) up to order $q$, implies the appearance of terms of order $q^2$ in $\Gamma^{(1)}_{M-Q}$ as shown in (\ref{(26)}). However such
terms have not been included in (\ref{(30.2)}) since they do not play any role in the discussion below. Indeed, in the case $y=1$ (along the symmetry axis) all these terms vanish, and
in the case
$y\neq1$, in the limit $R->2$, the leading term among the factors multiplying $q^2$, is of the same order of magnitude as that multiplying $q$, therefore since we are considering small
values of
$q$, all terms with $q^2$ may be neglected.

In  the next section we shall analyze the motion of a test particle in a space--time described by the M-Q$^{(1)}$ solution.
\section{The geodesics for the  M-Q$^{(1)}$ metric}
The equations governing the geodesics can be derived from the Lagrangian
\begin{equation}
2{\cal L}=g_{\alpha\beta}\dot{x}^\alpha\dot{x}^\beta,
\end{equation}
where the dot denotes differentiation with respect to an affine parameter
$s$ , which for timelike geodesics coincides with the proper time. Then,
from the Euler-Lagrange equations,
\begin{equation}
\frac{d}{ds}\left(\frac{\partial{\cal
L}}{\partial\dot{x}^\alpha}\right)-\frac{\partial{\cal L}}
{\partial x^\alpha}=0,
\end{equation}

it follows:
\begin{equation}
 \ddot t g_{tt}+\dot t(\dot r g_{tt,r}+g_{tt,\theta} \dot \theta)=0
\label{theta}
\end{equation}
\begin{equation}
2\ddot r g_{rr}+2\dot r(\dot r g_{rr,r}+g_{rr,\theta} \dot \theta)-\dot t^2 g_{tt,r}-\dot r^2 g_{rr,r}-\dot \theta^2 g_{\theta \theta,r}-\dot \phi^2 g_{\phi \phi,r}=0
\label{ere}
\end{equation}
\begin{equation}
2\ddot \theta g_{\theta \theta}+2\dot \theta(\dot r g_{\theta \theta,r}+g_{\theta \theta,\theta} \dot \theta)-\dot t^2 g_{tt,\theta}-\dot r^2 g_{rr,\theta}-\dot \theta^2 g_{\theta
\theta,\theta}-\dot
\phi^2 g_{\phi
\phi,\theta}=0
\label{teta}
\end{equation}
\begin{equation}
\ddot \phi g_{\phi \phi}+\dot \phi(\dot r g_{\phi\phi,r}+g_{\phi \phi,\theta} \dot \theta)=0
\label{fi}
\end{equation}
Where, our exterior metric in Erez--Rosen coordinates read:
\begin{eqnarray}
g_{rr} &  = & -e^{2 \Gamma-2 \Psi} \left(1+\frac{\lambda^2\sin^2\theta}{1-2
\lambda}\right)=  -e^{2q(\Gamma_{q^1}-
\Psi_{q^1})}\frac{1}{\left(1-\frac{2M}{r}\right)}\nonumber\\
\vspace{0.2cm}
g_{\theta \theta} &  = &- e^{2 \Gamma-2 \Psi} r^2 (1-2 \lambda +\lambda^2
\sin^2\theta) = -r^2 e^{2q(\Gamma_{q^1}- \Psi_{q^1})} \nonumber\\
\vspace{0.2cm}
g_{\phi\phi} &  = & -e^{-2 \Psi} r^2 \sin^2 \theta(1-2 \lambda) = -r^2 \sin^2
\theta  \ e^{-2q\Psi_{q^1}}\nonumber\\
\vspace{0.2cm}
g_{tt} &  = &  e^{2 \Psi} = \left(1-\frac{2M}{r}\right)\  e^{2q\Psi_{q^1}}
\label{(31)}
\end{eqnarray}
and  the functions  $\Gamma_{q^1}$ and
$\Psi_{q^1}$
are given by
\begin{eqnarray}
\Psi_{q^1} &  = & \frac{5}{32} (3y^2-1)(3 R^2-6
R+2)\ln\left(1-\frac2R\right) \nonumber\\
\vspace{0.2cm}
&-&\frac58 \ln\left(1-\frac2R\right)-\frac54 
\frac{R-1}{(R-1)^2-y^2}+\frac{15}{16} (3 y^2-1) (R-1) \quad ,
\label{(30.1N)}
\end{eqnarray}
\begin{eqnarray}
\Gamma_{q^1} &  = &-\frac{15}{8} 
(1-y^2)(R-1)\ln\left(1-\frac2R\right) \nonumber\\
\vspace{0.2cm}
&-&\frac{15}{4} (1-y^2)-\frac54 (1-y^2)
\left[\frac{(R-1)^2+y^2}{((R-1)^2-y^2)^2}\right]
\label{(30.2N)}
\end{eqnarray}
 Since we are concerned only with timelike geodesics,  the range of our coordinates is:
$$\infty >t\geq 0  \qquad r> 2M  \qquad \pi \geq \theta \geq 0  \qquad 2\pi \geq \phi \geq 0$$

\subsection{Radial geodesics}

Let us consider the motion of a test particle along a radial geodesic, for an arbitrary value of $\theta$. Thus putting $\dot \theta=\dot \phi=0$ in (\ref{ere}), we obtain 
\begin{equation}
2 \ddot r g_{rr}+\dot r^2(g_{rr,r}+\frac{g_{rr}g_{tt,r}}{g_{tt}})-\frac{g_{tt,r}}{g_{tt}}=0
\label{radial1}
\end{equation}

or using (\ref{(31)})
\begin{equation}
\ddot r=-e^{2q(\Psi_{q^1}-\Gamma_{q^1})}(\frac{M}{r^2}+\frac{q(R-2)}{R}\Psi_{q^1,r})-\dot r^2 q \Gamma_{q^1,r}
\label{radial2}
\end{equation}
with $\Psi_{q^1}, \Gamma_{q^1}, \Psi_{q^1,r}$ and $\Gamma_{q^1,r}$
easily obtained from (\ref{(30.1N)}, \ref{(30.2N)}).

In the spherically symmetric case ($q=0$) we obtain from (\ref{radial2}) the well known result
\begin{equation}
\ddot r=-\frac{M}{r^2}
\label{radial5}
\end{equation}

Now, before proceeding farther, and in order to express our results in terms of physically meaningful quantitites, let us introduce
(locally defined) coordinates associated with a locally Minkowskian observer (or alternatively a tetrad field associated with such Minkowskian observer). Thus, let
\begin{equation}
dX=\sqrt{-g_{rr}}dr
\label{x}
\end{equation}
and
\begin{equation}
dT=\sqrt{g_{tt}}dt
\label{t}
\end{equation}
from where it follows that
\begin{equation}
\dot r=\frac{\frac{dX}{dT}}{\sqrt{g_{rr}\left[(\frac{dX}{dT})^2-1\right]}}
\label{radial6}
\end{equation}
and
\begin{equation}
\frac{d^2 X}{dT^2}=\ddot r\sqrt{-g_{rr}}\left[1-(\frac{dX}{dT})^2\right]^2-(\frac{dX}{dT})^2 \frac{g_{rr,r}\left[1-(\frac{dX}{dT})^2\right]}{2(-g_{rr})^(3/2)}
\label{radial7}
\end{equation}
In the spherically symmetric case ($q=0$), (\ref{radial7}) reduces to
\begin{equation}
\frac{d^2 X}{dT^2}=-\frac{M}{r^2}\left[1-(\frac{dX}{dT})^2\right](1-\frac{2M}{r})^{-1/2},
\label{ss}
\end{equation}
or, in terms of $R$
\begin{equation}
\frac{d^2 X}{dT^2}=-\frac{R,_{r}}{R^{3/2}}\left[1-(\frac{dX}{dT})^2\right](R-2)^{-1/2},
\label{ssR}
\end{equation}
which is also a known result  \cite{Mc}. Since $(\frac{dX}{dT})$ is always smaller than one, the attractive nature of gravity for any value of $r$ (larger than $2M$) is clearly
exhibited in (\ref{ss}).

Let us now see what is the situation for any value of $q$, arbitrarily small but different from zero, as we approach the horizon. We  shall analyze two different situations, namely:
\begin{itemize}
\item The particle moves along a radial geodesics, outside the symmetry axis ($y^2\neq1$)
\item The particle moves along the symmetry axis ($y^2=1$)
\end{itemize}

\subsubsection{$y^2\neq1$}
Let us assume that the test particle moves along a radial geodesics outside the symmetry axis, then  as $R->2$, the leading term in 
 (\ref{radial7}), for small values of $q$ is 
\begin{equation}
\frac{d^2 X}{dT^2}\longrightarrow \frac{-R,_{r}\left[1-(\frac{dX}{dT})^2\right]}{2^{\frac{3}{2}}(R-2)^{1/2}}+\frac{{\cal O} (q)}{(R-2)^{1/2}}
\label{limit3}
\end{equation}
 Observe that this limit (except for small corrections
of order
$q$) is the same as the one obtained in the exactly spherically symmetric situation  from (\ref{ssR}). Thus for particles moving along radial geodesics (excluding the symmetry axis),
small values of the quadrupole moment introduce small perturbations on the trajectories, and the attractive nature of the gravitational force is preserved at all  times. Let
us now see how this situation change for particles moving along the symmetry axis ($y^2=1$).

\subsubsection{$y^2=1$}

In this case, it follows that as $R->2$,
\begin{equation}
\ddot r\longrightarrow -q\frac{5}{16}\frac{R,r}{(R-2)}e^{-\frac{5q}{4(R-2)}},
\label{limit}
\end{equation}
which is of course a positive quantity if $q<0$, and:
\begin{equation}
\sqrt{-g_{rr}}\longrightarrow \frac{\sqrt{2}}{\sqrt{R-2}}e^{\frac{5q}{8(R-2)}}
\label{limit1}
\end{equation}
and
\begin{equation}
g_{rr,r}\longrightarrow \frac{5qR,_{r}}{2(R-2)^3}e^{\frac{5q}{4(R-2)}}
\label{limit2}
\end{equation}

Using (\ref{limit}-\ref{limit2}) in (\ref{radial7}), we finally find that as $R->2$
\begin{equation}
\frac{d^2 X}{dT^2}\longrightarrow \frac{-5qR,_{r}}{2^{\frac{7}{2}}(R-2)^{3/2}}\left[1-(\frac{dX}{dT})^2\right]e^{\frac{-5q}{8(R-2)}}
\label{limit3}
\end{equation}
which is a positive quantity if $q<0$, diverging at the horizon, for any, different from zero, value of $q$. Thus, small deviations
form spherical symmetry (with
$q<0$) produce, close to the horizon, a positive (in the sense of increasing $X$) radial acceleration of particles moving radially on the symmetry axis, as measured by a Lorentzian
observer. Therefore, under such circumstances,  such observer would infer the existence of a ``repulsive'' gravitational force acting on the particle. This conclusion is valid in
all locally Minkowskian frames, including of course the instantaneous rest frame where $\frac{dX}{dT}=0$. Thus for example, if a test particle,initially  at rest with respect to the
Minkowskian  local frame, is placed close to the horizon (on the axis of  symmetry), as seen by such observer, it will tend to displace in the sense of increasing $X$, moving away from
the source. 

If $q>0$ then
$\frac{d^2 X}{dT^2}$ will be negative and tending to zero as
$r$ approaches
$2M$.

It is worth noting that the source of this difference between the $y^2=1$ and the $y^2\neq 1$ cases is the the term $\Psi_{q^1,r}$. Indeed, in
the later case ($y^2\neq 1$) we have , as $R->2$,
\begin{equation}
\Psi_{q^1,r} \longrightarrow -\frac{15 R,_{r}(1-y^2)}{16(R-2)},
\label{radial4IIII}
\end{equation}
 whereas in the case $y^2=1$, we have 
\begin{equation}
\Psi_{q^1,r} \longrightarrow \frac{5 R,_{r}}{8(R-2)^2},
\label{radial4III}
\end{equation}

It should be emphasized, once again, that the results above (\ref{limit}-\ref{limit3}), hold  for any small, but {\em finite} value of $q$. Therefore one should not expect to obtain the
spherically symmmetric limit (\ref{ssR}) from  (\ref{limit3}) as
$q\rightarrow 0$, since the spherically symmmetric contribution (in the limit $R\rightarrow 2$) being smaller, has been omitted in (\ref{limit3}).

Finally, observe that the fulfilment of equation (\ref{teta}), imposes a  constraint on $\dot r$ which does not play any role in the very appearance of the repulsive effect as implied
by equation (\ref{limit3}) since, as already mentioned, such an effect is observed by all locally Minkowskian observers, including the instantaneously at rest  observer. 
\subsection{Circular geodesics}
Let us now consider circular geodesics. From (\ref{theta})-(\ref{fi}), and  the definition of the angular velocity of a test particle along circular geodesics
\begin{equation}
\omega=\frac{\dot \phi}{\dot t}
\label{omega}
\end{equation}
we obtain using $\dot r=\dot \theta=0$, in the equatorial plane ($\theta=\pi/2$)

\begin{equation}
\omega^2=\frac{e^{4q\Psi_{q^1}}\left[R,_{r}+Rq\Psi_{q^1,r} (R-2)\right]}{M^2 R^3(R,_{r}-qR\Psi_{q^1,r})}
\label{omegaI}
\end{equation}

In the spherically symmetric case ($q=0$), (\ref{omegaI}) yields the Kepler law
\begin{equation}
\omega^2=\frac{M}{r^3}
\label{kepler}
\end{equation}

However, for arbitrarily small (different from zero) quadrupole moment, (\ref{omegaI}) exhibits a bifurcation from Kepler law, as $R->2$. Nevertheless, such bifurcation takes place for
orbits for which the tangential velocity of the particle is larger than 1, and therefore are deprived of physical interest (see below).
Indeed, we have  for the tangential velocity $W^{\mu}$ of the test particle along the circular geodesic (see for example \cite{Anderson})
\begin{equation}
W^{\alpha}=\frac{d\phi}{\sqrt{g_{tt}}dt}\delta^{\alpha}_{\phi}
\label{tang1}
\end{equation}

Then,  we obtain on the equatorial plane

\begin{equation}
W^2=\frac{R,_{r}+Rq\Psi_{q^1,r}(R-2)}{(R-2)(R,_{r}-Rq\Psi_{q^1,r})}
\label{Wsnsph}
\end{equation}

It follows from (\ref{Wsnsph}) that photonic orbits ($W=1$) differ from $r=3M (R=3)$ by factors of order $q$. Also, if $R->2$, then
$W^2->\frac{8}{15q}$. Of course these orbits are clearly unphysical, being $W>1$.

Finally it is worth noting that circular geodesics on the equatorial plane are stable. Indeed, it follows from (\ref{teta}) with $\dot r=\dot \theta=0$ that 
\begin{equation}
2\ddot\theta g_{\theta \theta}=\dot t^2(g_{tt,\theta}+\omega^2 g_{\phi \phi,\theta})
\label{estabili}
\end{equation}
implying $\ddot \theta=0$ on the equatorial plane.
\section{Conclusions}
We have first argued that the M-Q$^{(1)}$ solution represents a good candidate among known Weyl solutions, to describe small deviations from spherical symmetry. Then we have calculated
the geodesic equations for a test particle moving in that space--time.

We have found that in general the behaviour of circular geodesics is not very different from the spherically symmetric case. However important differences apppear in the behaviour of
radial geodesics. Thus, it has been established that particles along the axis of symmetry, close to $R=2$ feel a repulsive force if $q<0$ (corresponding to an oblate source). It should
be obvious that such an effect is absent in Newtonian theory, for small ($q<<1$) values of the quadrupole moment. 

It goes without saying  that we have not obtained the full description of the
trajectories of test particles in the M-Q$^{(1)}$ spacetime, since we have not completely integrated the geodesic equations. Certainly such integration would be necessary in order to
extract definitive conclusions about the behaviour of the test particle. Unfortunately we were not able to find analytical solutions to this problem, and a numerical approach  is
 out of the scope of this paper.

 It is worth mentioning that repulsives forces in the context of general relativity have been brought out before, the best known example probably being the case in the
Reissner--Nordstrom solution (see for example \cite{papapetrou}). Also, they appear  in connection with rotating and/or unbounded sources
\cite{repulsiveI} or as spurious phenomena associated with coordinate effects (see
\cite{Mc} and references therein).

As an example of the later is the bouncing back of particles that strike the surface r=2M, found by Rosen \cite{Rosen} for the spherically symmetric case. However observe that the
proper radial acceleration as measured by a  locally Minkowskian observer, for the spherically symmetric case, does not exhibit a change of sign in the limit $r->2M$, as indicated by
(\ref{ss}).

In our case, the repulsive character of the force (understood as a change of sign in $\frac{d^2 X}{dT^2}$) is exhibited by means of a quantity measured by  locally Minkowskian
observers, and therefore it is not a coordinate effect, and takes place close to, but before reaching the surface $r=2M$. In this sense the following comment is in order: It should be
clear that at $R=2$ there are not locally Minkowskian observers, since that is a singular surface for any non--vanishing value of $q$  \cite{quevedo2}. However, for sufficiently small
(but finite) values of $q$, close to but at a ``finite'' distance from $2M$ (where locally Minkowskian observers do exist) the leading term in $\frac{d^2 X}{dT^2}$ is given by
(\ref{limit3}). In other words it is in principle possible to find a region where locally Minkowkian observers exist and $\frac{d^2 X}{dT^2}$ change of sign.

Finally it is worth asking what are the physical implications of the study presented here. In view of the above, we would say that any result implying the presence of strong
gravitational fields and spherical symmetry, should be tested against small fluctuations of such symmetry. Also, it remains to be seen which, if any,  might be  the
implications of the reported repulsive effect, on some astrophysical scenarios, as for example in relativistic jets \cite{Mashhoon}.

\end{document}